\documentclass[preprint]{aastex}

\shorttitle{Luminosities and Space Densities of GRBs}
\shortauthors{Schmidt}

\begin{document}

\title{Luminosities and Space Densities of Gamma-Ray Bursts}

\author{Maarten Schmidt}
\affil{California Institute of Technology, Pasadena, CA 91125}
\email{mxs@deimos.caltech.edu}

\begin{abstract}

We use a homogeneous sample of gamma-ray bursts (GRB) extracted from
5.9 years of BATSE DISCLA data \citep{sch99} and a variety of assumed
broken power-law luminosity functions to derive GRB luminosities and 
space densities. Luminosity functions that are narrow or exhibit no 
density evolution produce expected redshift distributions that are 
incompatible with the observation of a GRB redshift of 3.4. 
For $q_o = 0.1$ and density evolution rising to 10 
at $z=1$, we find for a variety of slopes of the luminosity function
values of the local space density around 0.18 Gpc$^{-3}$ y$^{-1}$
with a range of only $40\%$. Characteristic $50-300$ keV peak luminosities 
exhibit a range of a factor of 6 around
$6 \times 10^{51}$ erg s$^{-1}$, corresponding to a characteristic total
luminosity of $1.2 \times 10^{53}$ erg in the $10-1000$ keV band. For
$q_o = 0.5$, densities are higher and luminosities lower, both by a
factor of 2.5. The local emissivity of GRBs is
$1.0 \times 10^{52}$ erg Gpc$^{-3}$ y$^{-1}$ in the $10-1000$ keV band.

\end{abstract}


\section{Introduction}

Soon after the launch of the {\it Compton Gamma Ray Observatory} in 1991,
observations of gamma-ray bursts (GRB) with the Burst and Transient
Source Experiment (BATSE) established two important global properties
of GRBs: an isotropic sky distribution \citep{mea92} and a 
radial distribution incompatible with a uniform population in
euclidean space \citep{meb92}. The evidence for a non-uniform 
radial distribution was based on either
the $N(>P)$ distribution or the $V/V_{max}$ distribution of the observed
GRBs. The $N(>P)$ distribution of the peak flux $P$ should be a power
law of slope $-3/2$ for a uniform population in euclidean space. The
observed distribution was shallower, particularly at lower
fluxes \citep{mea92}. For a uniform population, the $V/V_{max}$ 
distribution would be uniform between 0 and 1, with $<V/V_{max}> = 0.5$ 
\citep{sch88}. Early BATSE observations yielded $<V/V_{max}> = 0.35$,
significantly different from the value expected for homogeneity 
\citep{meb92}. 

A cosmological distance scale of GRBs is compatible with the isotropic
sky distribution and the radial distribution \citep{pac92}. This has
been confirmed by the first observation of a large redshift for a GRB
afterglow \citep{met97}. The distance scale in this cosmological 
scenario for GRBs has been derived from the observed $N(>P)$ distribution, 
usually on the assumption that GRBs are standard candles 
\citep{pen96,wij98,tot99}. \citet{kru98} have shown that relaxing the
standard candle assumption renders a broad range of models consistent 
with the BATSE $N(>P)$ relation.  

In this {\it Letter}, we are reporting the results of setting the
cosmological distance scale of GRBs, assuming a full luminosity function
and using the euclidean $<V/V_{max}>$ value as a distance 
indicator. We use the BD2 sample of GRBs derived from 5.9 years
of BATSE DISCLA data, briefly described in Sec. 2.
The characterization of the luminosity function and its evolution, and
the derivation of predicted distributions of luminosity, $V/V_{max}$,
redshift, and flux are discussed in Sec. 3. The results for a variety of
luminosity functions are presented in Sec. 4 and the conclusions are 
summarized in Sec. 5. We are assuming in this paper a Hubble constant 
$H_o = 70$ km s$^{-1}$ Mpc$^{-1}$ and zero cosmological constant.

\section{The BD2 Sample of Gamma-Ray Bursts}

We use a large homogeneous sample (the BD2 sample) of 1391 GRBs derived 
from BATSE DISCLA data, consisting of the continuous data stream from
the eight BATSE detectors in four energy channels on a time scale of
1024 msec \citep{fis89}. 
This sample is a revision, described below, of the BD1 sample
which resulted from a search of DISCLA data over the period TJD 
$8365-10528$ \citep{sch99}. The BD1 sample was based on a trigger 
algorithm that used the background both before and after the onset of 
the burst and required an excess of at least $5\sigma$ over background 
in at least two detectors in the energy range $50-300$ keV. 

In the process of classifying triggers for the creation of the 
BD1 sample \citep{sch99}, we had accepted 1018 DISCLA triggers that were 
within 230 sec of a GRB in the BATSE catalog \citep{mee99} as genuine GRBs. 
In addition, we classified another 404 DISCLA triggers as GRBs. 
In a subsequent revision,
we have now inspected the output of the BATSE detectors over a time interval 
of 12,000 sec around each of the 404 GRBs not in the BATSE catalog, as
well as those in the catalog for which times or positions differed 
appreciably. In the process, we rejected 7 triggers as caused by source 
fluctuations, 18 turned out to be parts of other GRBs of long duration, 
and 6 were identified as the soft repeater SGR1806-20 and rejected. 
As a consequence of the revision, the BD2 
sample now contains 1391 GRBs, of which 1013 are listed in the BATSE catalog, 
another 377 are classified as GRBs, and one as a probable GRB. 

For the purpose of this paper, we can characterize the BD2 sample
as follows. The number of GRBs is 1391. The sample effectively represents
2.003 years of full sky coverage \citep{sch99}, so the rate is 694 GRBs 
per year. The average euclidean $V/V_{max}$ is $0.334\pm0.008$. 
The limiting flux has a distribution $G(P_{lim})$ that has been derived 
as follows. For 104 positions of fixed celestial coordinates distributed
isotropically around the sky, we checked every 100 sec during every 
tenth day whether the sky position was above the horizon and whether the 
time did not fall in an exclusion window set up to avoid interference 
or bad data \citep{sch99}. If so, we converted the limiting count of 5 
times the square root of the observed background in the second brightest 
illuminated detector into a limiting flux $P_{lim}$. 
The resulting distribution $G(P_{lim})$ has a median at 
0.37 ph cm$^{-2}$ s$^{-1}$, 
and 10 and 90 percentiles at 0.29 ph cm$^{-2}$ s$^{-1}$ and 
0.51 ph cm$^{-2}$ s$^{-1}$, respectively, in the $50-300$ keV band. 

\section{Derivation of the Distance Scale}

Since the number of redshifts of GRBs is as yet too small for a
statistical derivation of the luminosity function, we will instead 
{\it assume} a full luminosity function and then use the observed
number of GRBs and their euclidean $<V/V_{max}>$ in the BD2 sample 
to derive properties such as the local space density and
the characteristic luminosity $L^*$. 
We will find that these properties vary relatively little as we
change the shape of the luminosity function.

\subsection{The Luminosity Function}

Experience has shown that the differential luminosity function of many types 
of extragalactic objects can be represented as a broken power law
(cf. e.g. \citet{has98} for X-ray active galactic nuclei). We will
use a broken power law, and in addition introduce upper and lower
limits of luminosity, thus allowing both narrow (standard candle) and more
realistic broad luminosity functions. We will also assume density
evolution, to be introduced below.

The local luminosity function of peak GRB luminosities $L$, defined 
as the co-moving space density of GRBs in the interval $\log L$ to
$\log L + d\log L$, is
\begin{mathletters}
\begin{eqnarray}
\Phi_o(L) & = & 0, \qquad\hbox{for}\qquad \log L < \log L^* - \Delta_1, \\
\Phi_o(L) & = & c_o (L/L^*)^{\alpha_1}, \qquad\hbox{for}\qquad
 \log L^* - \Delta_1 < \log L < \log L^*, \\
\Phi_o(L) & = & c_o (L/L^*)^{\alpha_2}, \qquad\hbox{for}\qquad
 \log L^* < \log L < \log L^* + \Delta_2, \\
\Phi_o(L) & = & 0, \qquad\hbox{for}\qquad \log L > \log L^* + \Delta_2.
\end{eqnarray}
\end{mathletters}

Given that most types of extragalactic objects show evolution, and considering
that the rate of star formation may be relevant for GRBs \citep{mad98}, 
we introduce density evolution $\rho(z)$. In most cases, we have assumed
\begin{eqnarray}
\rho(z) & = & (1+z)^p, \qquad\hbox{for}\qquad 0 < z < z_p, \\ 
\rho(z) & = & (1+z_p)^p, \qquad\hbox{for}\qquad z > z_p. 
\end{eqnarray}
While any type of evolution can be explored \citep{che99},
we have used in this {\it Letter}
only the case $z_p = 1.0$ and $\rho(1.0) = 10$ without a decline toward
larger redshifts \citep{ste99}. The luminosity function at redshift $z$ 
is $\Phi_z(L) = \Phi_o(L) \rho(z)$.

\subsection{Predicting the Sample Properties from the Luminosity Function}

In principle, with the full characterization of the BD2 sample 
given in Sec. 2 and the luminosity function $\Phi_z(L)$ with all its free 
parameters set, we can predict or model all properties of the sample. 
Our procedure will be to assume values for the parameters $\Delta_1$,
$\Delta_2$, $\alpha_1$, and $\alpha_2$, and then to vary $L^*$ until the
predicted value of the euclidean $<V/V_{max}>$ equals the observed value
in the BD2 sample, and to set $c_o$ to fit the observed number of GRBs.

The modeling procedure involves the derivation of the peak flux $P(L,z)$ 
of a GRB of peak luminosity $L$ observed at redshift $z$ \citep{sch86},
\begin{equation}
P(L,z) = {L\over 4\pi A^2(z)}{C(E_1(1+z),E_2(1+z))\over C(E_1,E_2)}
\end{equation}
where $A(z)$ is the bolometric luminosity distance and $C(E_1,E_2)$
is the integral of the spectral energy distribution
between $E_1 = 50$ keV and $E_2 = 300$ keV. We use a Band spectrum
\citep{ban93} for the energy distribution with $\alpha = -1.0$,
$\beta = -2.0$, and $E_o = 200$ keV in the restframe.

Objects with luminosity $L$ observed in a part of the BD2 sample
with flux limit $P_{lim}$ are detectable to a maximum redshift 
$z_{max}(L,P_{lim})$ that is easily derived from eq.(4).
The total number of objects in the sample is
\begin{equation}
N_{BD2} = \int \Phi_o(L) \, d\log L \int G(P_{lim}) \, dP_{lim} 
\int^{z_{max}(L,P_{lim})}_0 \rho(z)\, (dV/dz)\, dz\
\end{equation}
Based on this model we can derive the distributions of the peak flux $P$,
the euclidean $V/V_{max} = (P/P_{lim})^{-3/2}$, the peak luminosity $L$,
and the redshift $z$. These form the basis for the discussion of results
below.

\placetable{tbl-1}

\section{Results for a Variety of Luminosity Functions}

We exhibit in Table~\ref{tbl-1} numerical results for 18 luminosity function
models, nine each for the cases $q_o = 0.1$ and $q_o = 0.5$. 
All are based on a Hubble constant $H_o = 70$ km s$^{-1}$ Mpc$^{-1}$
and zero cosmological constant.
Among the nine cases, two have zero density evolution, six have evolution
as described in eqs. (2) and (3), and one is an exponential of cosmic
time. In the six cases, we explored variations of the extent of the 
luminosity function below $L^*$, as well as of the logarithmic slopes 
of both parts, cf. Table~\ref{tbl-1}.

\placefigure{fig1}

As an example, model 14 uses the deceleration parameter $q_o = 0.1$ and 
logarithmic slopes of $-0.5$ and $-2.0$ for the power law luminosity 
function, which extends from $L^*/10$ to $100\,L^*$. The model has 
density evolution $\rho(z) = (1+z)^{3.32}$ for $0 < z < 1$ and 
$\rho(z) = 10.0$ for $z > 1$. Figure~\ref{fig1} shows the relation between the 
model values of $<V/V_{max}>$ and $\log L^*$. The observed value of 
$<V/V_{max}>=0.334 \pm 0.008$ corresponds to $\log L^* = 51.76 \pm 0.08$, 
where the error is the formal error corresponding to that in $<V/V_{max}>$. 
The value of $c_o$ is set by the total number of GRBs in the BD2 sample.
With the luminosity function of model 14 fully characterized, we
can derive the predicted distributions in the BD2 sample of the
luminosities, the fluxes, the $V/V_{max}$ values and the redshifts; the last 
three are illustrated in Figure~\ref{fig2} together with the luminosity 
function. Apart from the obvious onset of incompleteness in the observed
fluxes, the predicted $N(>P)$ agrees well with the observed distribution. 
The model distribution of $V/V_{max}$ shows fairly good agreement
with the observations. The distribution of redshifts
is broad, with some redshifts expected as large as 6. The median redshift
is 1.5. 

\placefigure{fig2}

Besides the values of $\log L^*$ and $c_o$ derived for each model, we show
in Table~\ref{tbl-1} also the local ($z=0$) space density $\rho_o$ of GRBs 
integrated 
over the full range of luminosities $L$, and the probability $P_{3.4}$
that a GRB in the BD2 sample has a redshift $z>3.4$. This probability
is of interest, since one of the first observed GRB redshifts was 3.4 
\citep{kul98}. For the models with a standard candle luminosity function 
or with zero evolution (nos. $11-13$ and $51-53$), the values of $P_{3.4}$ 
are so low that they are inconsistent with the observed redshift of 3.4.
Model 53 is similar to the standard candle luminosity models considered 
by \citet{wij98} and \citet{tot99}. \citet{wij98} derived a luminosity 
about 3.5 times larger, and a local density 2.5 times smaller than 
those given in Table 1, while \citet{tot99} found a luminosity  
similar to our value. The broad luminosity functions considered by
\citet{kru98} cover a range from $10^{50} - 10^{52}$ erg s$^{-1}$,
overlapping with those in models $54 - 58$. 

We show in Figure~\ref{fig3} a plot of $\rho_o$ vs. $L^*$ for all models.
There are clear systematics that can be summarized as follows. From
$q_o = 0.1$ to $q_o = 0.5$ local densities increase and characteristic
luminosities decrease, both by a factor of about 2.5. The effect of the
density evolution adopted is to decrease local densities by a factor of
$15-20$ and to increase $L^*$ by a factor of 3. For given cosmology and
evolution, local densities range over a factor less than 2 and
$L^*$ over a factor of 6. 
Given that $\rho_o$ and $L^*$ tend to be inversely correlated, we
derived for each model $E_{out}$, the total peak output of GRBs 
per unit volume at $z=0$.
As shown in Table~\ref{tbl-1}, $E_{out}$ is remarkably stable at 
$5 \times 10^{50}$ erg s$^{-1} $Gpc$^{-3}$ y$^{-1}$ for the adopted evolution.
For zero evolution, $E_{out}$ is some 5 times larger.

\placefigure{fig3}

If the gamma radiation of GRBs is beamed or collimated, say over 
$4\,\pi\,F$ steradians, then all luminosities should be multiplied by $F$,
and all space densities divided by $F$. The total peak output $E_{out}$ 
is unaffected by beaming.

\section{Summary}

We have shown that using primarily $<V/V_{max}>$, we can derive properties
of GRBs of useful accuracy. We have used redshift information only
in arguing that the observation of a redshift of 3.4 for a GRB 
is unlikely if the luminosity function is a standard candle, or if the GRB
rate does not evolve with redshift. For density evolution rising to a factor
of 10 at $z=1$ and $q_o = 0.1$ (for $q_o = 0.5$, cf. Table~\ref{tbl-1}),
the local space density is 0.18 Gpc$^{-3}$ y$^{-1}$, with a range of only
40\% among the models considered. 
Characteristic peak luminosities range over a factor of 6 around
$6 \times 10^{51}$ erg s$^{-1}$ in the $50-300$ keV band. 

Luminosities\footnote{An investigation of the reason why our flux limits 
$G(P_{lim}$ (cf. end of Sec. 2) are higher than those given by
\citet{pen98} suggests the possibility that all our fluxes $P$ may be
high by around $27\%$ compared to BATSE fluxes. If this is confirmed,
then the values of $\log L^*$ in Table 1 should be corrected by $-0.10$ 
and the emissivities $E_{out}$ reduced by $27\%$, while space densities
are essentially unchanged.} 
derived in this study are peak luminosities (strictly per 1024 msec), 
since the detection of GRBs in the BD2 sample is on that time scale. We
estimate the {\it total} energy radiated by GRBs in gamma rays by 
integrating over the time profile and by extending the spectral range to
$10-1000$ keV. To account for the total energy radiated in the 
$50-300$ keV band integrated over the duration of the burst, we consulted
data in the BATSE catalog \citep{mee99}. We find that for the $100-200$ 
strongest bursts, the ratio of fluence over peak flux is around $9-10$ s.
The ratio of the energy radiated in a band of $10-1000$ keV
over that in the $50-300$ keV range based on the Band spectrum 
\citep{ban93} with $\alpha = -1.0$, $\beta = -2.0$, and $E_o = 200$ keV
is 2.1. We conclude that total luminosities over the
$10-1000$ keV band are 20 times larger than the $50-300$ keV peak
luminosities. Employing this factor, we obtain for $q_o = 0.1$ a
characteristic GRB luminosity of $1.2 \times 10^{53}$ erg in the $10-1000$
keV band. Similarly, the local emissivity of GRBs $E_{out}$ discussed in
Sec. 4 is $1.0 \times 10^{52}$ erg Gpc$^{-3}$ y$^{-1}$ in the $10-1000$ keV 
band with very little range among the models.

\clearpage

\figcaption[schmidtfig1.ps]{Plot of the euclidean value of $<V/V_{max}>$ 
versus peak luminosity for luminosity function model 14, cf. 
Table~\ref{tbl-1}. \label{fig1}}

\figcaption[schmidtfig2.ps]{Properties of luminosity function model 14, cf. 
Table~\ref{tbl-1}. 
Panel (a) shows the luminosity function $\Phi(L)$ (Gpc$^{-3}$ y$^{-1}$
per unit $\log L$). The break luminosity $L^*$
has been set by $V/V_{max}$ and the normalization by the observed
rate of GRBs. Panel (b) compares the model euclidean $V/V_{max}$ 
distribution for the BD2 sample with the observed distribution. 
Panel (c) compares the model flux distribution $N(>P)$ for the BD2 sample
with the observations. Panel (d) shows the expected 
distribution of redshifts for the BD2 sample, with an indication of 
the probability $P_{3.4}$ of a redshift exceeding 3.4. \label{fig2}}  

\figcaption[schmidtfig3.ps]{Plot of the local GRB density versus peak 
luminosity $L^*$ for 18 luminosity function models, cf. Table~\ref{tbl-1}. 
The systematic effects of the cosmological model ($q_o=0.1$ and 
$q_o=0.5$) and of evolution are easily discerned. For a given 
cosmological model and evolution, local densities range over 
a factor of 2, and luminosities over a factor of 6 for the different
luminosity function models considered. \label{fig3}}


\vskip 0.5truein

\begin{deluxetable}{ccccccrccrlrrr}
\footnotesize
\tablecaption{Properties of various luminosity functions models. \label{tbl-1}}
\tablewidth{0pt}
\tablehead{
\colhead{\#} & \colhead{$q_o$\tablenotemark{a}} & 
\colhead{$p$}  & \colhead{$z_p$}  &
\colhead{$\Delta_1$} &
\colhead{$\Delta_2$}  & \colhead{$\alpha_1$} & \colhead{$\alpha_2$} &
\colhead{$\log L^*$\tablenotemark{b}} & 
\colhead{$c_o$\tablenotemark{c}}     & 
\colhead{$\rho_o$\tablenotemark{d}} &
\colhead{$P_{3.4}$\tablenotemark{e}}   &
\colhead{$E_{out}$\tablenotemark{f}}
}
\startdata
$11$&0.1&0.00&0.0&$-$0.1&0.1&$-$0.5&$-$2.0&51.11&10.2 &2.0 & 0\%&26.1 \\
$12$&0.1&0.00&0.0&$-$1.0&2.0&$-$0.5&$-$2.0&51.34&1.30 &2.7 & 1\%&29.0 \\
$13$&0.1&3.32&1.0&$-$0.1&0.1&$-$0.5&$-$2.0&51.44&0.654 &0.13& 0\%&3.6 \\
$14$&0.1&3.32&1.0&$-$1.0&2.0&$-$0.5&$-$2.0&51.76&0.0740&0.15& 5\%&4.4 \\
$15$&0.1&3.32&1.0&$-$2.0&2.0&$-$0.5&$-$2.0&52.20&0.0276&0.22&13\%&5.3 \\
$16$&0.1&3.32&1.0&$-$1.0&2.0&   0.0&$-$2.0&51.62&0.128&0.16& 5\%&4.3 \\
$17$&0.1&3.32&1.0&$-$1.0&2.0&$-$1.0&$-$2.0&51.92&0.0364&0.15& 5\%&4.3 \\
$18$&0.1&3.32&1.0&$-$1.0&2.0&$-$0.5&$-$1.0&51.43&0.0842&0.19&14\%&5.8 \\
$19$&0.1&4.04\tablenotemark{g}&1.0&$-$1.0&2.0&$-$0.5&$-$2.0&51.72&0.0780&0.16& 5\%&4.2 \\
$51$&0.5&0.00&0.0&$-$0.1&0.1&$-$0.5&$-$2.0&50.71&25.0 &5.0 & 0\%&25.7 \\
$52$&0.5&0.00&0.0&$-$1.0&2.0&$-$0.5&$-$2.0&50.96&3.18 &6.6 & 1\%&29.6 \\
$53$&0.5&3.32&1.0&$-$0.1&0.1&$-$0.5&$-$2.0&51.08&1.66 &0.33& 0\%& 4.0 \\
$54$&0.5&3.32&1.0&$-$1.0&2.0&$-$0.5&$-$2.0&51.42&0.189&0.39& 5\%& 5.0 \\
$55$&0.5&3.32&1.0&$-$2.0&2.0&$-$0.5&$-$2.0&51.77&0.0857&0.69& 9\%&6.1 \\
$56$&0.5&3.32&1.0&$-$1.0&2.0&   0.0&$-$2.0&51.26&0.334&0.40& 4\%&4.9 \\
$57$&0.5&3.32&1.0&$-$1.0&2.0&$-$1.0&$-$2.0&51.57&0.0927&0.38& 5\%&5.0 \\
$58$&0.5&3.32&1.0&$-$1.0&2.0&$-$0.5&$-$1.0&51.06&0.234&0.54&11\%&7.0 \\
$59$&0.5&3.56\tablenotemark{g}&1.0&$-$1.0&2.0&$-$0.5&$-$2.0&51.34&0.205&0.43& 4\%&4.6 \\
\enddata

\tablenotetext{a}{In all cases we use $H_o = 70$ km s$^{-1}$ Mpc$^{-1}$
and zero cosmological constant.}
\tablenotetext{b}{$L^*$ is the peak luminosity in erg s$^{-1}$ in the
$50-300$ keV band, if the GRB is radiating isotropically.}
\tablenotetext{c}{Units are Gpc$^{-3}$ y$^{-1}$ per unit $\log L$.}
\tablenotetext{d}{$\rho_o$ is the local ($z=0$) GRB rate, in units of
Gpc$^{-3}$ y$^{-1}$.}
\tablenotetext{e}{Probability of $z>3.4$ for a GRB in the BD2 sample.}
\tablenotetext{f}{Local peak energy output of GRBs in the $50-300$ keV
band, in units of 10$^{50}$ erg s$^{-1}$ Gpc$^{-3}$ y$^{-1}$.} 
\tablenotetext{g}{In this case $\rho(z) = e^{p\, \tau(z)}$ where $\tau(z)$
is the light travel time, expressed in the age of the universe.}

\end{deluxetable}

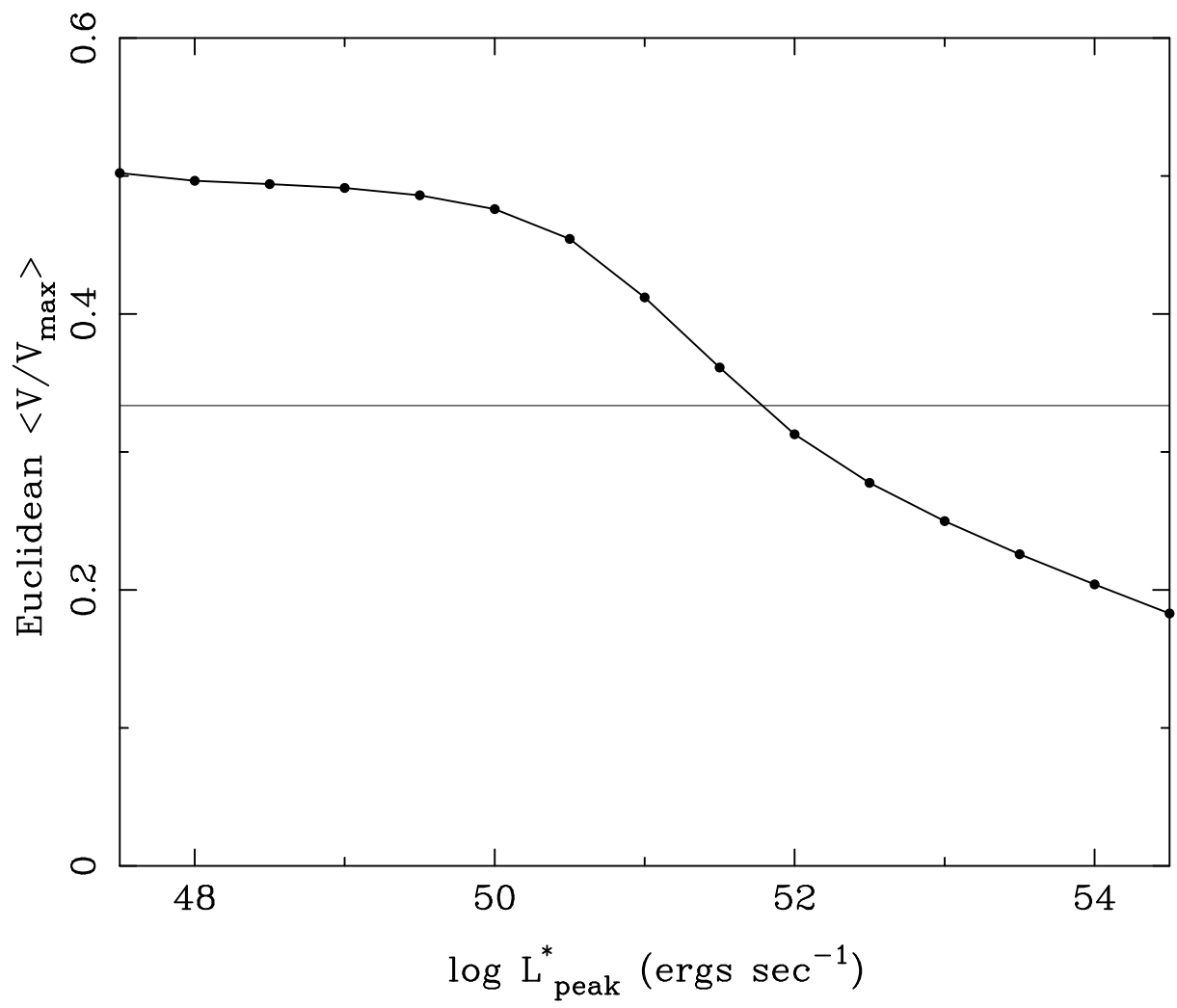
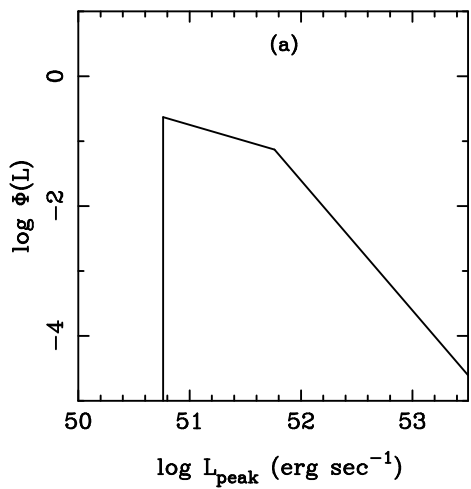
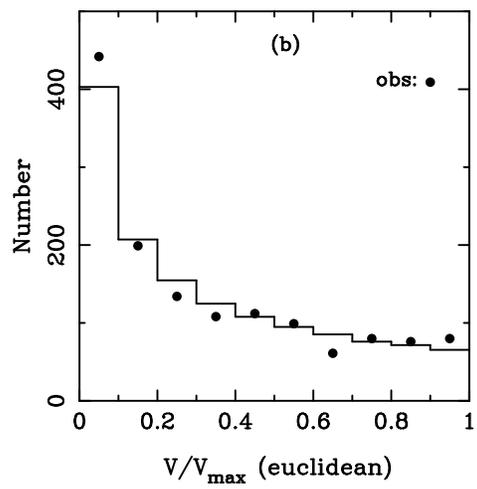
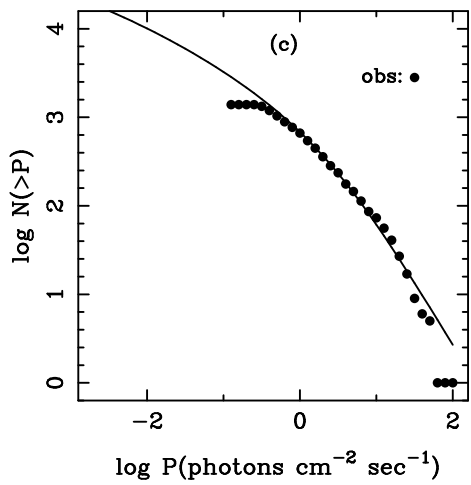
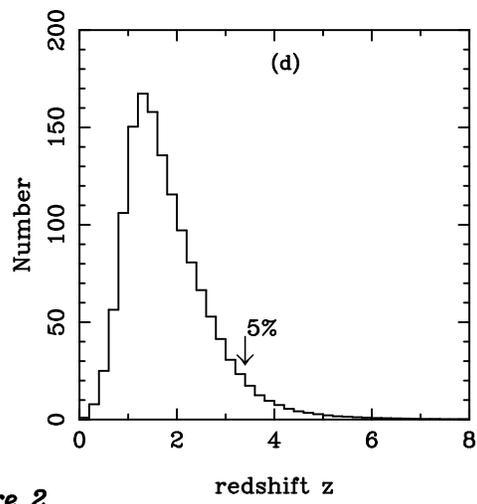
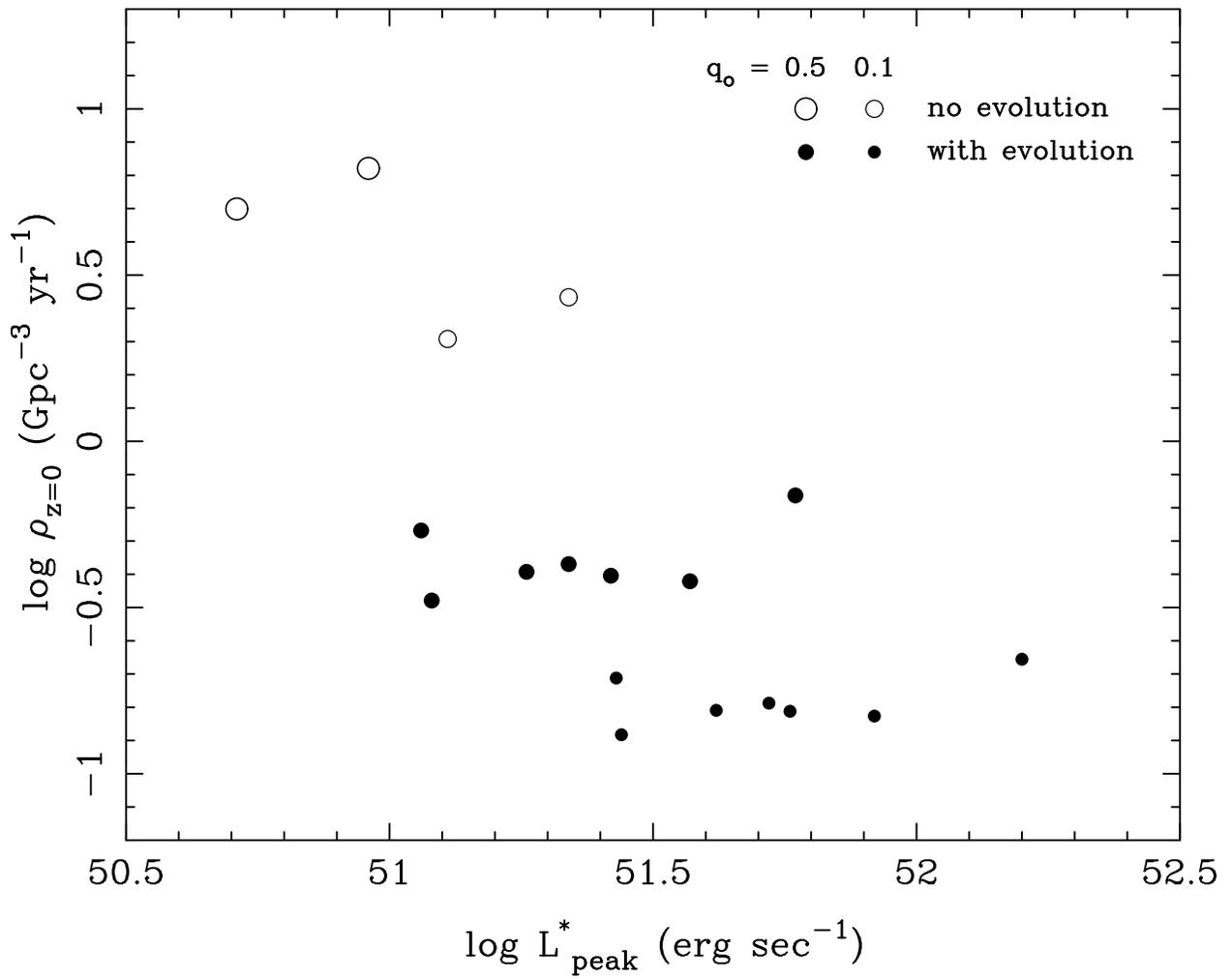
\end{document}